 \documentclass[final,2p,times,twocolumn]{elsarticle}

\usepackage{amssymb}
\usepackage[english]{babel}
\usepackage[utf8]{inputenc}
\usepackage{graphicx}
\usepackage{lineno}
\usepackage{textcomp, gensymb}
\usepackage{nomencl}
\makenomenclature
\usepackage{fullpage}
\usepackage{breqn}
\usepackage{amsmath}
\usepackage{booktabs}
\usepackage{float}
\usepackage{subfig}
\usepackage{xcolor}
\usepackage{newunicodechar}
\usepackage{adjustbox}
\usepackage{cuted}%

\usepackage{overpic}
\usepackage[flushleft]{threeparttable}
    \usepackage{lineno,hyperref}
    \modulolinenumbers[5]
    \usepackage{multirow}
    \biboptions{numbers,sort&compress}

\journal{TBD}

\begin{document}

\begin{frontmatter}



\title{Phase-Field Model of Silicon Carbide Growth During Isothermal Condition}


\author[inst1]{Elias J. Munoz}
\author[inst1]{Vahid Attari}
\author[inst1]{Marco Martinez}
\author[inst2]{Matthew B. Dickerson}
\author[inst1]{Miladin Radovic}
\author[inst1]{Raymundo Arroyave}

\affiliation[inst1]{organization={Department of Materials Science Engineering, Texas A\&M University},
            city={College Station},
            state={TX},
            country={USA}}

\affiliation[inst2]{organization={Materials and Manufacturing Directorate, Air Force Research Laboratory},
            city={Wright-Patterson AFB},
            state={OH},
            country={USA}}

\begin{abstract}
\par Silicon carbide (SiC) emerges as a promising ceramic material for high-temperature structural applications, especially within the aerospace sector. The utilization of SiC-based ceramic matrix composites (CMCs) instead of superalloys in components like engine shrouds, combustors, and nozzles offers notable advantages, including a 25\% improvement in fuel efficiency, over 10\% enhanced thrust, and the capability to withstand up to 500 $^{\circ}$C higher operating temperatures. Employing a CALPHAD-reinforced multi-phase-field model, our study delves into the evolution of the SiC layer under isothermal solidification conditions. By modeling the growth of SiC between liquid Si and solid C at 1450 $^{\circ}$C, we compared results with experimental microstructures and quantitatively examined the evolution of SiC thickness over time. Efficient sampling across the entire model space mitigated uncertainty in high-temperature kinetic parameters, allowing us to predict a range of growth rates and morphologies for the SiC layer. The model accounts for parameter uncertainty stemming from limited experimental knowledge and successfully predicts relevant morphologies for the system. Experimental results validated the kinetic parameters of the simulations, offering valuable insights and potential constraints on the reaction kinetics. We further explored the significance of multi-phase-field model parameters on two key outputs, and found that the diffusion coefficient of liquid Si emerges as the most crucial parameter significantly impacting the SiC average layer thickness and grain count over time. 
This study contributes insight into microstructure evolution in the Si-C binary system relevant to engineering CMCs in industry.
\end{abstract}




\end{frontmatter}

\section{Introduction}\label{S:1}
\par
\begin{figure*}[!ht]
    \centering
    \includegraphics[width=0.8\textwidth]{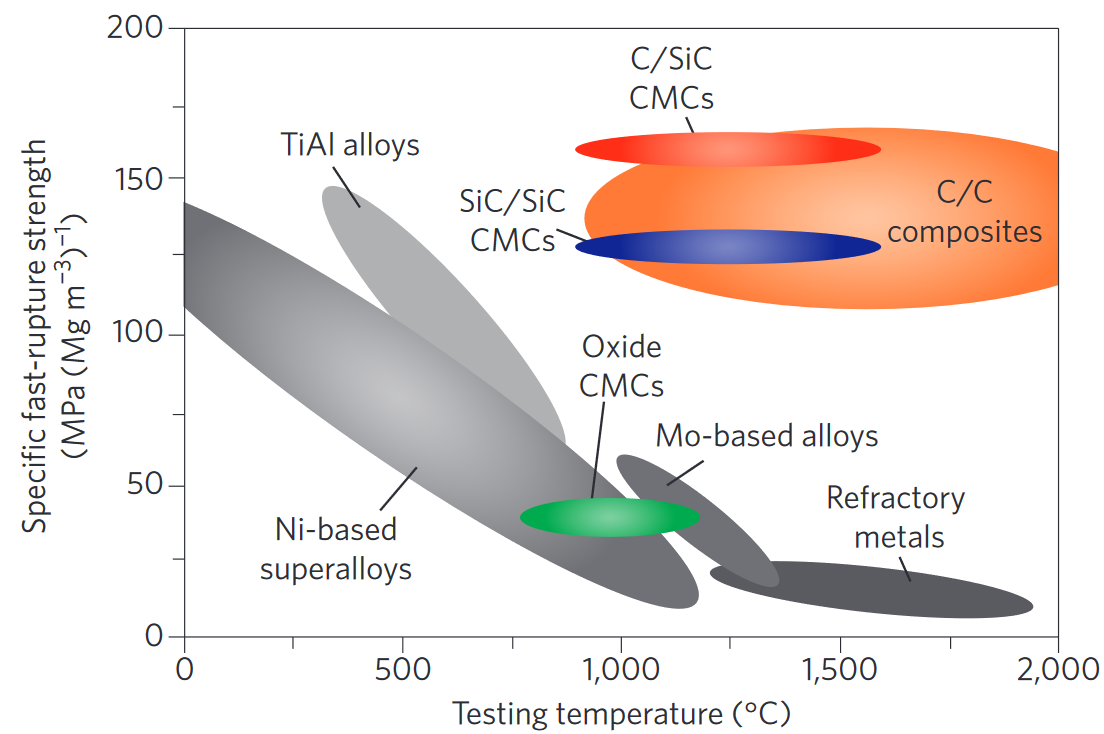}
    \caption{Comparison of various high-temperature structural materials strength and temperature data for ceramics and metal alloy~\cite{marshall_integral_2008}}
    \label{fig:MaterialvTemp}
\end{figure*}

 The aerospace industry faces increasing demands for enhanced performance metrics, such as prolonging part lifetime for cost savings and elevating turbine engine operating temperatures to improve fuel efficiency~\cite{maity2022high,padture_advanced_2016}. The progress in thermal barrier coating technology has surpassed the high-temperature capabilities of Ni-based superalloys, creating an opportunity to substitute metallic components with ceramic matrix composites (CMCs). As depicted in Figure \ref{fig:MaterialvTemp}, CMCs emerge as superior alternatives to current Ni-based alloys in terms of temperature resistance for aerospace applications~\cite{marshall_integral_2008}. Besides their elevated service temperature and reduced maintenance requirements, CMCs boast a lighter weight compared to superalloys (e.g., SiC/SiC CMCs exhibit 1/3 the density of Ni alloys), resulting in high specific strength~\cite{evans_physics_1994}. While extensive research on CMCs occurred in the 1980s and 1990s, processing challenges, prohibitive costs, and suboptimal performance had relegated the material to the sidelines until more recent advancements~\cite{marshall_integral_2008}.

Among the CMC materials, C/SiC and SiC/SiC systems are extensively utilized, offering appealing properties for aviation applications. Advanced SiC/SiC melt-infiltration (MI) CMCs are crafted by infusing molten Si into carbon and high-strength ceramic fiber-containing preforms~\cite{wang2004effect}. As the reaction initiates, molten Si undergoes a reaction with carbon to produce SiC. However, a portion of the infiltrated Si remains unreacted, resulting in a predominantly SiC matrix with a small residual Si content. Typically, the unreacted Si within the CMC matrix fills remaining porosity, imparting high interlaminar strength, excellent thermal conductivity, and hermeticity to shield SiC fibers from oxidation. Nevertheless, the presence of residual Si significantly constrains the temperature capabilities of SiC/SiC (e.g., Si melting temperature, $T_{m}$, is 1410 $^\circ$C, while the upper operating temperature of SiC under oxidizing conditions is ~1600 $^\circ$C). Despite the commercialization of the MI production process for SiC/SiC, there are still gaps in fundamental knowledge concerning the early stages of the MI process and subsequent SiC growth.

One issue that has attracted considerable experimental activity is exploring and addressing processing issues, such as the choking of pores during CMC production ~\cite{mahajan_sic-based_2020}. More recently, modeling  approaches have been leveraged to investigate possible solutions to pore choking and overall optimization of reaction that will improve SiC growth and other properties of this material ~\cite{sergi_surface_2015,de_jong_numerical_1996,steiner_impact_2020}. Challenges associated with modeling arise due to the need for more precise experimental data on under the various temperatures at which this reaction occurs In the literature, cited work at high temperature data with respect to diffusion exists ~\cite{hon_self-diffusion_1979,hon_self-diffusion_1980,hong_self-diffusion_1979,hong_self-diffusion_1981,ghoshtagore_self-diffusion_1966} as well as lower than melting temperature of Si exists. Regrettably, at the intermediate temperatures data is severely lacking. For example, we lack interfacial energies, grain boundary energies, and diffusion values to properly parameterize the simulations. 

One significant challenge that has received considerable experimental attention involves the exploration and mitigation of processing issues, such as pore choking during CMC production~\cite{mahajan_sic-based_2020}. Recent efforts have turned to modeling approaches to address pore choking and optimize the overall reaction, aiming to enhance SiC growth and other material properties~\cite{sergi_surface_2015,de_jong_numerical_1996,steiner_impact_2020}. However, modeling encounters challenges due to the requirement for precise data across various temperatures at which this reaction occurs. While high-temperature data related to diffusion is available in the literature~\cite{hon_self-diffusion_1979,hon_self-diffusion_1980,hong_self-diffusion_1979,hong_self-diffusion_1981,ghoshtagore_self-diffusion_1966}, there is a notable scarcity of data at intermediate temperatures. More specifically, information on interfacial energies, grain boundary energies, and diffusion values is lacking, hindering the proper understanding of reaction outcomes. In this context, Reitz et al~\cite{reitz_reactive_2017} studied the infiltration of reactive melts into porous graphite tubes using Si and Si-Y alloys and showed that the interface reaction controls the wetting characteristics. There was a bimodal distribution of SiC grain sizes, suggesting that a dissolution/reprecipitation process is the primary mechanism of SiC growth, with solid-state diffusion playing a role at a later stage. Sergi et al.~\cite{sergi_surface_2015} investigated reactive molten silicon infiltration into carbon preforms using experiments and simulations (Lattice-Boltzmann method). The authors show Si infiltration into C occurs in two stages where during the first one, the reaction takes place in the liquid until a thick SiC layer forms, while during the second stage, a uniform and compact layer is formed when the reactants diffuse into the solid phase through the SiC phase. Likewise, Ness et al.~\cite{ness_microstructural_1986} examined the microstructural evolution of reaction-bonded silicon carbide, focusing on nucleation and growth, interfaces, and impurities. kulik et al.~\cite{kulik_modeling_2004} developed a 1D isothermal model of chemical vapor infiltration (CVI) accounting for the convective processes governed by phase transitions and transversal mass exchange between the pore systems. The authors show a transverse mass exchange affects the temporal evolution of the concentration of species in different pore systems (pores formed by fibers in a bundle or a system of inter-bundle pores) during the infiltration process. The process temperature range in these studies usually ranges from 1450\textdegree C to 2000\textdegree C. 


Controlling melt infiltration chemistry and reaction kinetics is instrumental for the applications of reactive melt-infiltration processes~\cite{eom_processing_2013,zou_microstructural_2010,reitz_reactive_2017,ness_microstructural_1986,tong_reactive_2016}. In this paper, we have used phase-field modeling to investigate the kinetics of SiC formation and growth from molten silicon and carbon. A thermodynamically consistent phase-field model (PFM) provides a robust foundation for exploring the model parameter space and its impact on the physical behavior of simulated materials~\cite{chen2022classical}. Utilizing a thermodynamically-consistent model, we study (a) the kinetics of SiC growth after a uniform layer forms, and (b) how material parameters affect the observed morphology. The study garners a better understanding of the kinetics of interfacial SiC growth, post-nucleation. Herein, we strove to identify the range of parameters (e.g., diffusivity, interfacial energy, and mobilities) that dominantly contribute to the reaction and evolution of SiC. As measuring these key parameters is extremely difficult at the temperature used for melt infiltration, our model sheds light on missing parameter information. A complete analysis of the Si infiltration and SiC nucleation process is beyond the scope of a mean-field model such as phase-field method. 
The rest of this article is arranged as follows: Section~\ref{S:2} discusses our proposed methods; Section~\ref{sec:results} presents the results and their discussion; and Section~\ref{sec:conc} provides a brief conclusion. Additional details related to some of the results are provided in the appendix section.


\section{Methods}
\label{S:2}
\subsection{Multi-Phase-Field Model}
A multi-phase-field formalism based on the method of equal-diffusion potential in the interface of binary phases~\cite{kim1999phase} was used to study the evolution of the microstructure at an isothermal and isobaric state. The method is based on the previous implementation taken from Attari et al.~\cite{attari_phase_2016,attari_interfacial_2018} as implemented for Cu/Sn/Cu electronic interconnections in 3D integrated circuits. There is a set of conserved (0 $\leq$ c$_i$ $\leq$ 1) and non-conserved (0 $\leq$ $\phi_i$ $\leq$ 1) variables that describe the composition and spatial fraction of the existing phases over the microstructure domain ($\Omega$), respectively. Accordingly, the total free energy of the chemically heterogeneous material involving interfacial ($f^{int}$) and bulk interactions ($f^{bulk}$) is:
\begin{equation}
    F^{tot} = \int_\Omega \left[f^{int} + f^{bulk} \right]d\Omega 
\end{equation}
\begin{equation}
f^{bulk} = \Sigma_i \Sigma_{j>i} \frac{4 \sigma_{ij}}{\eta_{ij}}[-\frac{\eta^2_{ij}}{\pi^2}\nabla \phi_i\cdot\nabla \phi_j + |\phi_i \phi_j|]
\end{equation}
\begin{equation}
f^{int}  = \Sigma_i \phi_i f^0_i (c_i)
\end{equation}
\noindent
where $\sigma_{ij}$ is the interface energy coefficient, $\eta_{ij}$ is the interface width, and $|\phi_i\phi_j|$ is the double obstacle potential. $f^0_i$ is the free energy of the homogeneous phase \emph{i} and \emph{$c_i$} is its molar concentration. All concentrations in this study are based on the molar concentration of Si in each phase. Using the model of the total free energy as a function of the conserved and non-conserved field variables, the following forms of the kinetic equations (diffusion and phase-field) are presumed as the governing equations~\cite{attari_phase_2016,attari_interfacial_2018}:
\begin{equation}\label{eq:diff}
   \frac{\partial c}{\partial t} = \nabla \cdot [D({\phi_i}) \sum_{i=1}^{N} \phi_i \nabla c_i],
\end{equation}
\begin{equation}\label{eq:pfm}
    \frac{\partial\phi_i}{\partial t} = - \sum_{i\neq j} \frac{M_{ij}}{N_p} [\frac{\partial F^{tot}}{\partial\phi_i} - \frac{\partial F^{tot}}{\partial\phi_j}]
\end{equation}
\noindent
where $D$ is the diffusivity coefficient, $M_{ij}$ is the interface mobility, and $N_p$ is the number of coexisting phases at the neighboring grid points. The phase-field equation (Eq.~\ref{eq:pfm}) applies only in the interface where $\phi_i$ changes between 0 and 1. However, the diffusion equation (Eq.~\ref{eq:diff}) is solved in the entire domain. Interdiffusivity ($D$) is defined as a function of the phase-field order parameter to take into account the diffusivity in various features of the microstructure including grain boundaries, bulk phases, and interfaces. Using mass conservation and the equality of chemical potentials, the coexistence of the phases in the interface is defined by:
\begin{equation}
    c(x,t) = \sum_{i=1}^{N} \phi_i c_i
\end{equation}
\begin{equation}
    f^1_{c_1}[c_1(x,t)] = f^2_{c_2}[c_2(x,t)] = \cdots = f^N_{c_N}[c_N(x,t)]
\end{equation}
\noindent
where $f^N_{c_N}$ stands as the derivative of the free energy with respect to the composition of the phase N. 

\subsection{Thermodyanmic Description of Si-C Binary Alloy}

The free energy equations for solid C and liquid Si phases can be written in the form of a CALPHAD functional expression for a binary solution:
\begin{equation}
   f^{\alpha}(c,T) = \sum_i c_i^{\alpha} . ^0G^{\alpha}_i + RT \sum_i c_i \ln(c_i) + \sum_i \sum_{j>i} c_i c_j \sum_v {^{\nu}L_{ij}^{\alpha}} (c_i - c_j)^{\nu} 
\end{equation}

\noindent
where $R$ is the ideal gas constant, $T$ is temperature, $^0G^{\alpha}_i$ is the reference Gibbs energy of phase $\alpha$, $^{\nu}L^{0}_{ij}$ represents the excess binary interaction parameter that is dependent on $\nu$. Si-C system is modeled by ${\nu}=0$ indicates a regular thermodynamic solution. In this study, the reference Gibbs energy terms and binary interactions for solid and liquid phases are obtained from the SGTE elemental database from Gr\"{o}bner et al., Lacaze et al., and Olesinski et al ~\cite{grobner_thermodynamic_1996,dinsdale_sgte_1991,olesinski_csi_1984,lacaze_assessment_1991}. SiC was referred to from the thermodynamic assessment of Cupid et al.~\cite{cupid_thermodynamic_2007}:
\small

\begin{equation}
  \begin{aligned}
    ^0G^{\text{Sol}}_{Si} & = G^{\text{Sol}}_C(1-c_{Si}) + G^{\text{Sol}}_{Si}(c_{Si}) \\
     + & RT\big(c_{Si}\log(c_{Si}) + (1-c_{Si})\log(1-c_{Si}) \big)  \\
     + & c_{Si}(1-c_{Si}) ^0L^{\text{Sol}}_{ij}
  \end{aligned}
\end{equation}

\begin{equation}
  \begin{aligned}
    ^0G^{Liq}_{Si} = & G^{Liq}_C(1-c_{Si}) + G^{Liq}_{Si}(c_{Si}) \\
    + & RT(c_{Si}\log(c_{Si}) + (1-c_{Si})\log(1-c_{Si})) \\
    + & c_{Si}(1-c_{Si}) ^0L^{Liq}_{ij}
  \end{aligned}
\end{equation}

\begin{equation}
  \begin{aligned}
    ^0G_{\text{SiC}} = & -88584+271.1462T-41.279T\log(T) \\
    - & 0.0043T^2 + 0.8\times10^6T^{-1}+0.2\times10^{-6}T^3
  \end{aligned}
\end{equation}

\begin{figure}[!h]
    \centering
    \includegraphics[width=0.9\columnwidth]{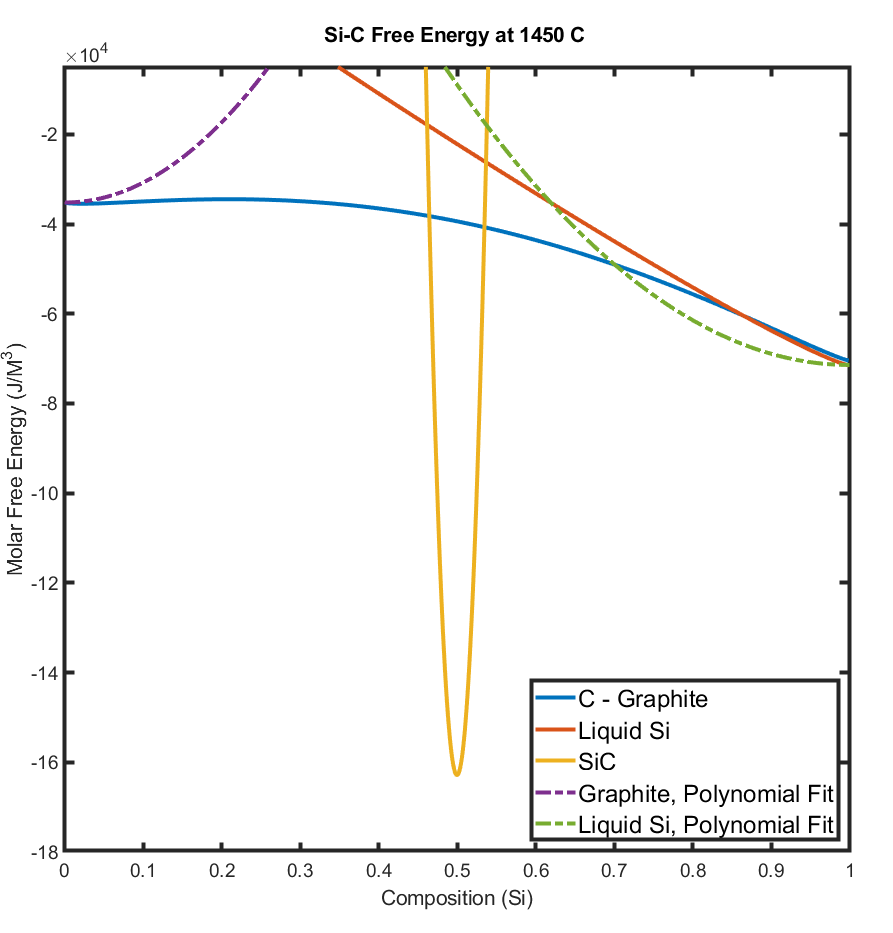}
    \caption{Free Energy curves for solid graphite, liquid silicon, and SiC IMC at 1450 $^{\circ}$C. Note that SiC is a stoichiometric phase and is shown as a line compound in the phase diagram. Here, we have assumed parabolic energy with a narrow composition range for ease of implementation.}
    \label{fig:Free_Energy}
\end{figure}

\normalsize
\noindent
where $c_{\text{Si}}$ is the concentration of silicon. The above CALPHAD equations enforce equilibrium concentrations of 0 and 1 for solid graphite and liquid silicon phases, respectively. With equilibrium concentrations close to zero, logarithmic potentials are difficult to work with making them unsuitable for use in the phase-field model. We substitute second-order polynomials for these potentials to model thermodynamic interactions between phases. The original CALPHAD forms of the free energy along with these adapted polynomials at 1450$^\circ$C are shown in Fig.~\ref{fig:Free_Energy}. The fitted energy functions for liquid silicon, solid graphite, and SiC IMC phases are respectively given as: 
\begin{equation}
f^{Liq}(c)=\rho_L(c_{Si}-1.0)^2 - 71593.43
\end{equation}
\begin{equation}
f^{Sol}(c)=\rho_S(c_{Si}-0.001)^2 - 33944.91
\end{equation}
\begin{equation}
f^{SiC}(c)=\rho_I(c_{Si}-0.50)^2 - 167363.17
\end{equation}

In these equations, $\rho$ is a fitting constant defined as $\rho_S=\rho_I=2\times10^6$ and $\rho_L=1\times10^4$ for solid, SiC, and liquid phases, respectively. 

\subsection{Simulation Setup and Parameters}

\par
\begin{figure}[h]
    \centering
    \includegraphics[width=0.9\columnwidth]{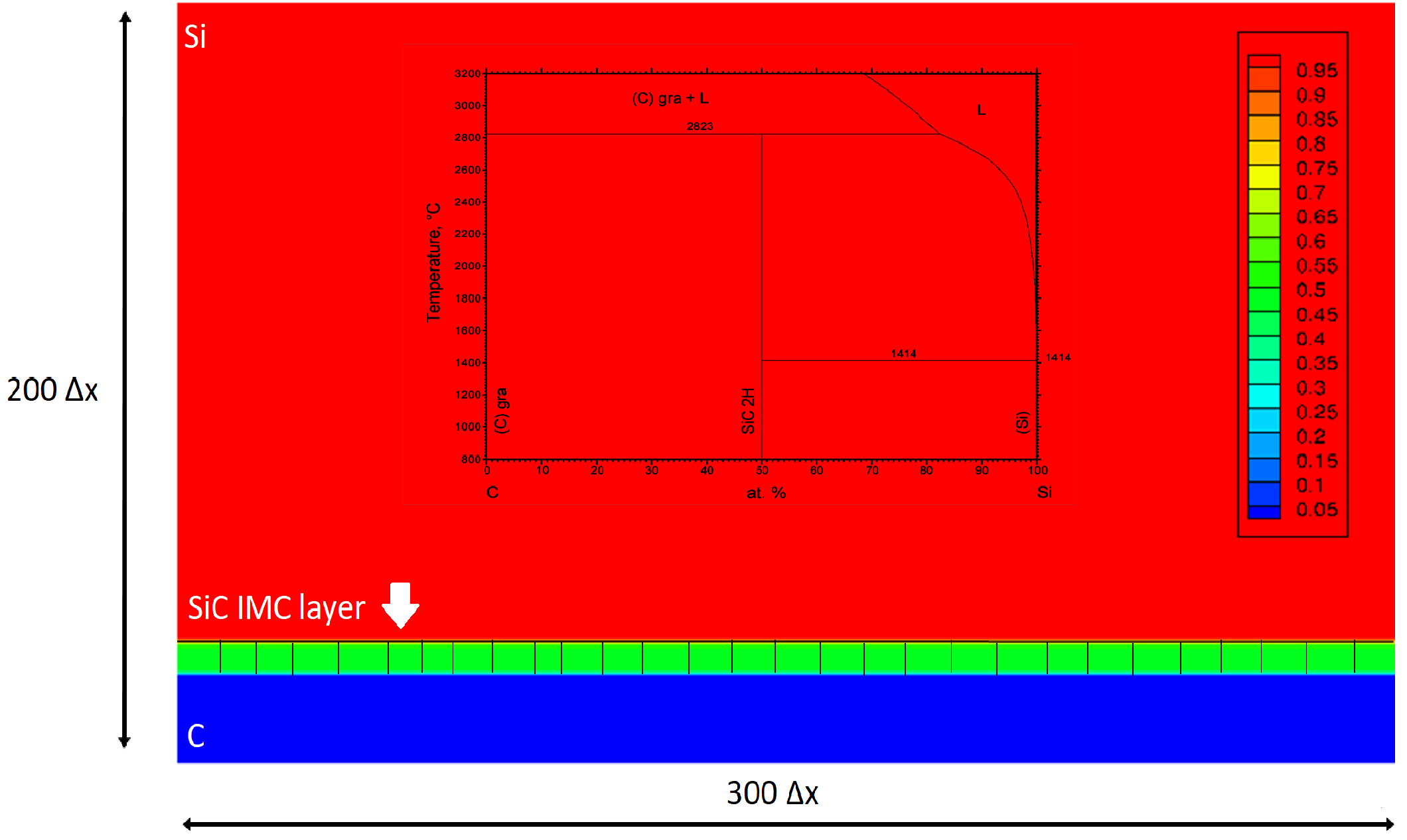}
    \caption{The schematic configuration of the simulation domain at the initialization. The legend and color map shows the concentration field and phase boundaries are shown by black solid lines. A phase diagram of SiC system is included in this plot. }
    \label{fig:schematic}
\end{figure}

 In this work, we used a 2-Dimensional (2D) domain for modeling the growth of SiC in the Si-C system at 1450 $^{\circ}$C. The simulation domain size is set to 300 ${\Delta}$x by 200 ${\Delta}y$ where ${\Delta}x={\Delta}y=0.25{\mu}m$ that corresponds to approximately 75$\times$30 microns. The width of the phase-field interface was set to be 4${\Delta}$x. The initial simulation domain is composed of three layered phases. The bottom layer consists of a pure carbon solid phase where a SiC layer is overlaid over it. The SiC layer is composed of rectangular SiC grains with the same height and different widths. The height of the SiC layer is $9{\Delta}y=2.25{\mu}m$. The rest of the simulation domain is filled with liquid silicon phase. We applied periodic boundary conditions at the right/left sides of the calculation domain and Neumann boundary conditions at the top/bottom sides for both composition and phase-field order parameters. The schematic configuration of the simulation domain is shown in Gif.~\ref{fig:schematic}.

\par The literature provides a set of prior values for diffusion, as outlined in Table~\ref{tab:Diffusion}. Due to the challenging experimental requirements, comprehensive investigations into Si-C kinetic parameters at high temperatures are limited. The existing data on diffusivities (lattice, grain boundary, and self-diffusion) served as our initial reference for simulations, as presented in Table~\ref{tab:Diffusion}. Since our simulations are conducted at lower temperatures, it is essential to reduce the diffusion values by several orders of magnitude. We extrapolated potential diffusion values based on Hong et al.'s trend line for bulk diffusion~\cite{hong_self-diffusion_1979} and interfacial diffusion values. Throughout the simulations, we iteratively adjusted the diffusivity parameters to attain optimum values, detailed in Table~\ref{tab:Kinetics}. Interfacial energy parameters and mobilities were adopted from the literature. Table~\ref{tab:Kinetics} provides a summary of all kinetic parameters along with upper and lower bounds for the parameter search. 

\begin{table}[!ht]
    \centering
\begin{tabular} { lcccc } 
 \toprule
 Parameter & $D_{0}$ ($cm^2/s$) & $E_{A}$ (eV) & Ref. \\ \hline
$D_{lc}$ & 2.62 \textpm $1.83\times10^{8}$ & 8.72 \textpm 0.14 & ~\cite{hon_self-diffusion_1979}\\
$D_{bc}$ & 4.44 \textpm $2.03\times10^{7}$ & 5.84 \textpm 0.09 & ~\cite{hon_self-diffusion_1979}\\
$D_{c}$ & 8.62 \textpm $2.01\times10^{5}$ & 7.41 \textpm 0.05 & ~\cite{hong_self-diffusion_1979}\\
$D_{Si}$ & 3.32 \textpm $1.43\times10^{7}$ & 8.20 \textpm 0.008 &
~\cite{hong_self-diffusion_1979}\\
$D_{Si}$ & 5.01 \textpm $1.71\times10^{2}$ & 7.22 \textpm 0.07 & 
~\cite{hong_self-diffusion_1979}\\
$D_{Si}$ & 1.54 \textpm $0.78\times10^{5}$ & 8.18 \textpm 0.10 & 
~\cite{hong_self-diffusion_1979}\\
$D_{bc}$ & 4.44 \textpm $2.03\times10^{7}$ & 5.84 \textpm 0.09 & 
~\cite{hong_self-diffusion_1979}\\
$D_{Si}$ & 8.36 \textpm $1.99\times10^{7}$ & 9.45 \textpm 0.05 & 
~\cite{hon_self-diffusion_1980}\\
$D_{C}$ & 8.62 \textpm $2.01\times10^{5}$ & 7.41 \textpm 0.05 & 
~\cite{hong_self-diffusion_1981}\\
\bottomrule
\end{tabular}
\caption{\label{tab:Diffusion} Diffusion parameters from literature where lc represents lattice diffusion coefficient, bc represents grain boundary diffusion coefficient, c represents self diffusion coefficient of \textsuperscript{14}C }   
\end{table}

\begin{table*}[!ht]
    \centering
\begin{threeparttable}
    \centering
    \caption{\label{tab:Kinetics} Kinetic parameters with the considered range. Note, parameters are defined in Table 2.}    
\begin{tabular} { llccc } 
 \toprule
 Variable Name & Phases & Parameter & Unit & Range \\ \hline
 Bulk diffusion & Si & $D_L$ & $cm^2/s$ & $5.0\times10^{-14}$ - $2.0\times10^{-12}$\\
  & C & $D_S$ & & $2.0\times10^{-20}$ - $5.0\times10^{-19}$\\
  & SiC & $D_I$ & & $4.0\times10^{-19}$ - $4.0\times10^{-18}$\\
 Interfacial diffusion & C/SiC & $D_{IS}$ & & $4.0\times10^{-16}$ - $1.0\times10^{-15}$ \\
  & Si/SiC & $D_{IL}$&  & $9.0\times10^{-16}$ - $4.0\times10^{-15}$ \\
 & SiC/SiC  & $D_{II}$ & & $4.0\times10^{-16}$ - $1.0\times10^{-14}$ \\
 Interface energy & C/SiC & $\sigma_{SI}$ & J/$m^2$  & 0.46 - 0.7 \\
   & SiC/SiC & $\sigma_{II}$ & & 0.27 - 0.40 \\
   & Si/SiC & $\sigma_{LI}$ & & 0.10 - 0.27  \\
 Interfacial Mobility & (Solid C and SiC) & $M_{SI}$ & $m^3/{J.s}$ & $9.02\times10^{-08}$ - $1.02\times10^{-07}$ \\
  & (Liquid and SiC) & $M_{IL}$ & & $1.0\times10^{-08}$ - $1.0\times10^{-06}$\\  \bottomrule
\end{tabular}
\begin{tablenotes}
      \item $^*$ S: Solid, L: Liquid, I: Intermetallic 
    \end{tablenotes}
\end{threeparttable}
\end{table*}


\subsection{Uncertainty Propagation and Model Parameter Importance Study}

Probabilistic uncertainty quantification/propagation (UQ/UP) methods have been extensively used to quantify the uncertainties of influential model parameters resulting from different aleatoric and epistemic sources and propagate them from model parameters to model outputs. 
Undertaking uncertainty propagation (UP) with computationally intensive models, such as meso-scale multi-physics phase-field models, can present a challenge. Propagating an ample number of samples from a desired distribution through such models is often impractical due to computational cost~\cite{attari2023towards}. Propagattion of the existing uncertainty in model inputs to model response (i.e., growth of SiC IMC layer), we require a large number of simulation runs where each run uses distinct combinations of parameters as the inputs. 

\par To propagate uncertainty in model parameters, we employ Saltelli sampling, implemented using the SALib Python package~\cite{Iwanaga2022,Herman2017}. Unlike traditional Monte Carlo sampling, which demands substantial computational resources, Saltelli sampling offers a more efficient alternative. This approach provides two methods to address the challenges of dimensionality in full-factorial design of experiments~\cite{saltelli_making_2002}~\cite{saltelli_variance_2010}. In the primary case, one can analyze, at a reduced computational cost, all effects of the first and total order, along with those of order $k-2$, at the expense of $n(k+2)$ simulations. The subsequent case requires a more significant number of function evaluations $n(2k+2)$ than the first case. This method estimates the indices of the first and total orders, plus all indices of order two and $k-2$. Although computationally more expensive, the second case provides deeper insights into pair interactions. We adopt the second method to generate multiple sample sets, including three interface energy parameters, six diffusivity parameters, and two mobility parameters, resulting in over 5000 samples. This allows us to obtain a dataset for analysis, identifying successful and failed simulations.

Subsequently, we leverage machine learning feature importance methods to assess the impact of the inputs of the multi-phase-field method on its outputs. Feature importance quantifies how much each input contributes to the model prediction, indicating the importance and utility of a specific variable. The phase-field data, obtained after a constant time of evolution, is divided into 30/70\% test and training data sets, and two machine learning feature importance models are constructed for analysis. Below is a brief explanation of these methods.

\subsubsection{Lasso Linear Regression Feature Importance}

To determine the importance of each input variable, we can fit a linear regression model and then extract the coefficients. The parameters of the model must be scaled before the fitting process can begin. For feature selection, Lasso regression with an L1 regularization parameter is used. In Lasso regression, the $\lambda$ regularization parameter controls the degree of regularization and shrinks the coefficients. Lasso can shrink correlated variables arbitrarily by selecting only one of the variables.

In linear models, the target value is modeled as a linear combination of features. Coefficients represent the relationship between the given feature $X_i$ and the target $y$, assuming that all the other features remain constant (conditional dependence). This is different from plotting $X_i$ versus $y$ and fitting a linear relationship: in that case, all possible values of the other features are taken into account in the estimation (marginal dependence).

\subsubsection{Permutation Feature Importance}

Permutation feature importance is a simple and commonly used technique that measures the increase in the model's prediction error after we permute the feature’s values, breaking the relationship between the feature and the true outcome. Permutations are applied to a Random Forest Regressor algorithm to achieve this. The permutations are performed on test data. 





\section{Results and Discussion}\label{sec:results}
\par Our simulations examine the effects of diffusivities, interfacial energies, and mobilities on SiC growth at an isothermal temperature of 1450$^\circ$C. We explored the morphology and rate of growth of the SiC layer by querying different combinations of the kinetic parameters as listed in Table~\ref{tab:Kinetics}. The simulations begin post-nucleation when a continuous SiC layer has already covered the C substrate. The composition of the three phases is set to equilibrium values using the Si-C phase diagram. As the simulation begins, the SiC layer consumes Si and C phases to promote SiC growth until it reaches the top of the simulation domain. The simulations account for three major diffusion pathways: bulk diffusion, interfacial diffusion, and grain boundary diffusion, and we aim to understand the dominant kinetics.

\par
\begin{figure*}[!ht]
    \centering
    \subfloat[]{\includegraphics[width=0.48\textwidth]{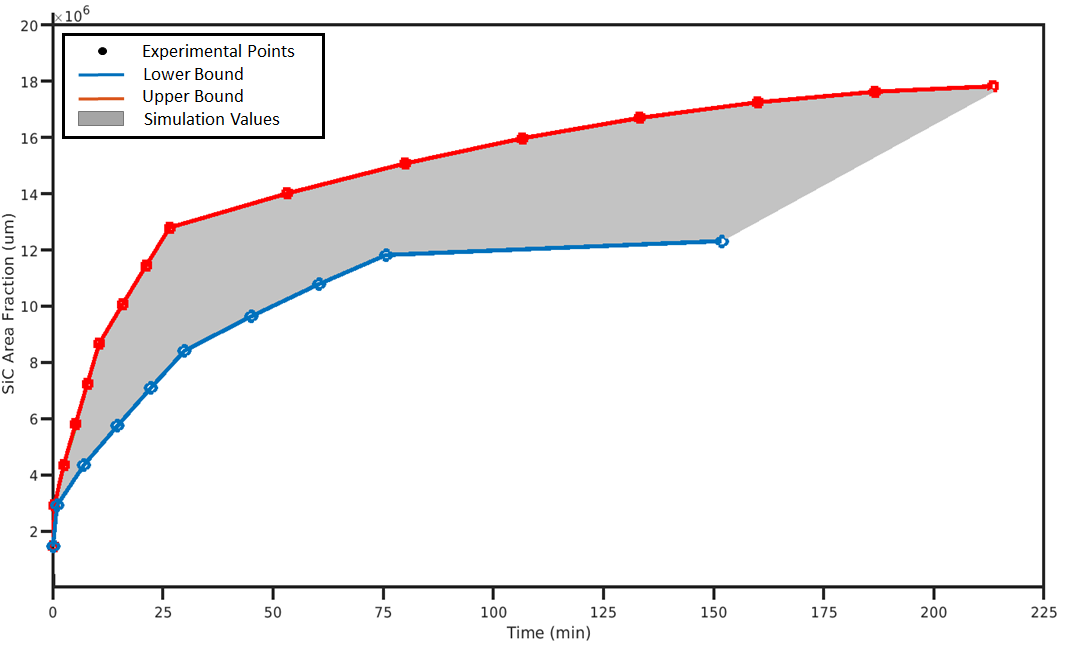}}
    \subfloat[]{\includegraphics[width=0.48\textwidth]{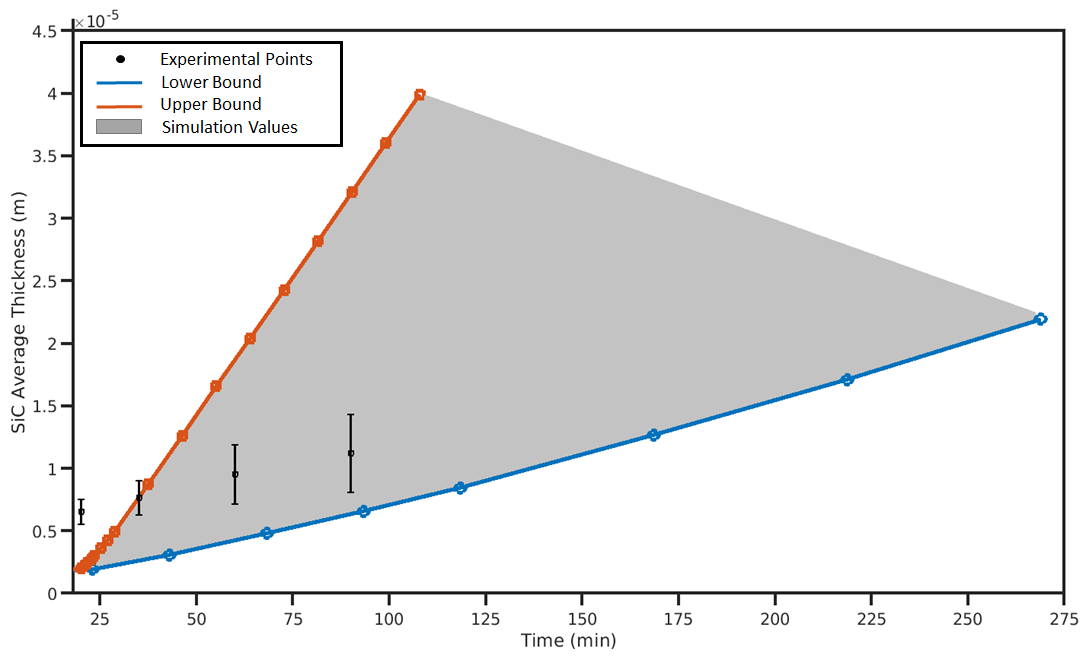}}
    \caption{Graphs for the growth of the SiC layer over time in 10,000 simulations. (a) Change of the area fraction of the SiC layer in time, (b) Change of the average thickness of the SiC layer in time. The upper and lower bounds in the simulations are shown in red and blue colored solid line and the intermediate calculations are shown in gray color. The experimental thickness and associated error bars from repeating the experiments are shown for thickness evaluations.}
    \label{fig:AvgThick}
\end{figure*}

 Figure~\ref{fig:AvgThick} provides a summary of two extracted quantities, namely the area fraction and average thickness of the SiC layer, over time from 10,000 simulations. The solid red/blue lines represent the upper and lower limits of these quantities, while the results from other simulations are depicted within the gray-colored area. In these simulations, the SiC layer experiences growth for approximately 250 minutes. The phase-field method predicts a parabolic growth for the area fraction of the SiC layer and a linear growth trend for thickness evolution. The black markers with error bars correspond to our experimental measurements at different annealing times (20, 40, 60, and 90 minutes).

The reactive melt infiltration experiments on graphite preforms with Si were isothermal, maintaining a temperature of 1450$^\circ$C for durations ranging from 2 minutes to 4 hours. SiC layer formation was investigated using a diffusion couple configuration. The samples, infiltrated at various times, were characterized by SEM to measure SiC growth. At 20 minutes, the initial SiC layer thickness slightly deviates from the simulation's initial setting, bearing implications for optimizing kinetic parameters based on experimental observations. 

\par We then analyze individual simulations or simulation groups, strategically identifying trends that may exist due to specific sets of parameters. In 5000 simulations, we systematically compared simulation groupings. This process provides crucial insights that direct the deliberate adjustment of parameter boundaries, bringing simulations closer to experimental outcomes. Furthermore, a detailed examination of simulation snapshots at various time markers provides insight into the kinetics, enabling a direct comparison of simulation thickness with experimental thickness. These snapshots serve as invaluable glimpses into the dynamics of phase evolution. We carefully interpret the kinetics, pinpointing the parameters exerting the most significant influence on simulation evolution. An illustrative example involves the identification of simulations successfully traversing multiple experimental points. By examining initial parameter conditions and iteratively refining constraints, we found simulation cases that align with an increasing number of experimental points.

\par
\begin{figure*}[!ht]
    \centering
    \begin{overpic}[width=0.99\textwidth]{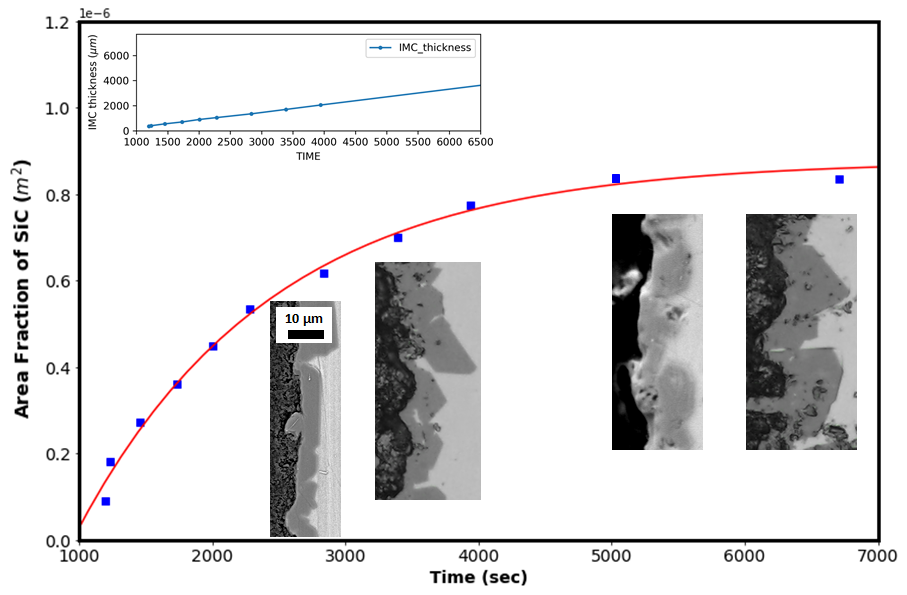}%
    \scriptsize\put(55,60){$A_f=-1.67e^{-06}(e^{-0.000681t}) + 8.77e^{-7}$}%
    \end{overpic}
    \caption{Area fraction of SiC layer as a function of time from a phase-field simulation. An inlay of the thickness of this layer over time is also shown in the top left corner of the graph. The experimental SEM micrographs are embedded in this graph and show the growth of the SiC layer.}
    \label{fig:AreaFraction}
\end{figure*}

 Figure~\ref{fig:AreaFraction} displays the outcome of an individual simulation, specifically the area fraction of the SiC layer. SEM images capturing the evolution of the SiC layer at different time intervals accompany this simulation result. The corresponding thickness result is presented in the top-left corner of the figure. The marker points represent extracted area fraction values for this simulation, and a solid line is obtained through exponential curve fitting (i.e., $y = ae^{bx} + c$). The functional form of this fit is outlined below: 


\begin{equation}
    A_f=-1.678e^{-06}(e^{-.0006815t}) + 8.771e^{-07}
\end{equation}



\par Figure~\ref{fig:AreaFraction} illustrates that the thickness of the SiC layer shows a linear progression over the simulation time. The blue line denotes the maximum thickness measured at each point along the line. Notably, the maximum thickness exhibits less variability compared to experimental instances. Additionally, our simulation thickness trend generally appears steeper than that observed in experimental results. This discrepancy could stem from continuous nucleation in experiments, while our simulations exclusively involve coarsening in the phase field model. Another possible factor is that the interface may not be fully resolved with the presence of SiC, leaving empty areas where nucleation can persist, accentuating the thickness difference in simulations. This effort established a range for kinetic parameters and presented existing uncertainties in defining a realistic description of the reactive melt infiltration process. 

\par
\begin{figure*}[!ht]
    \centering
    \begin{overpic}[width=0.99\textwidth]{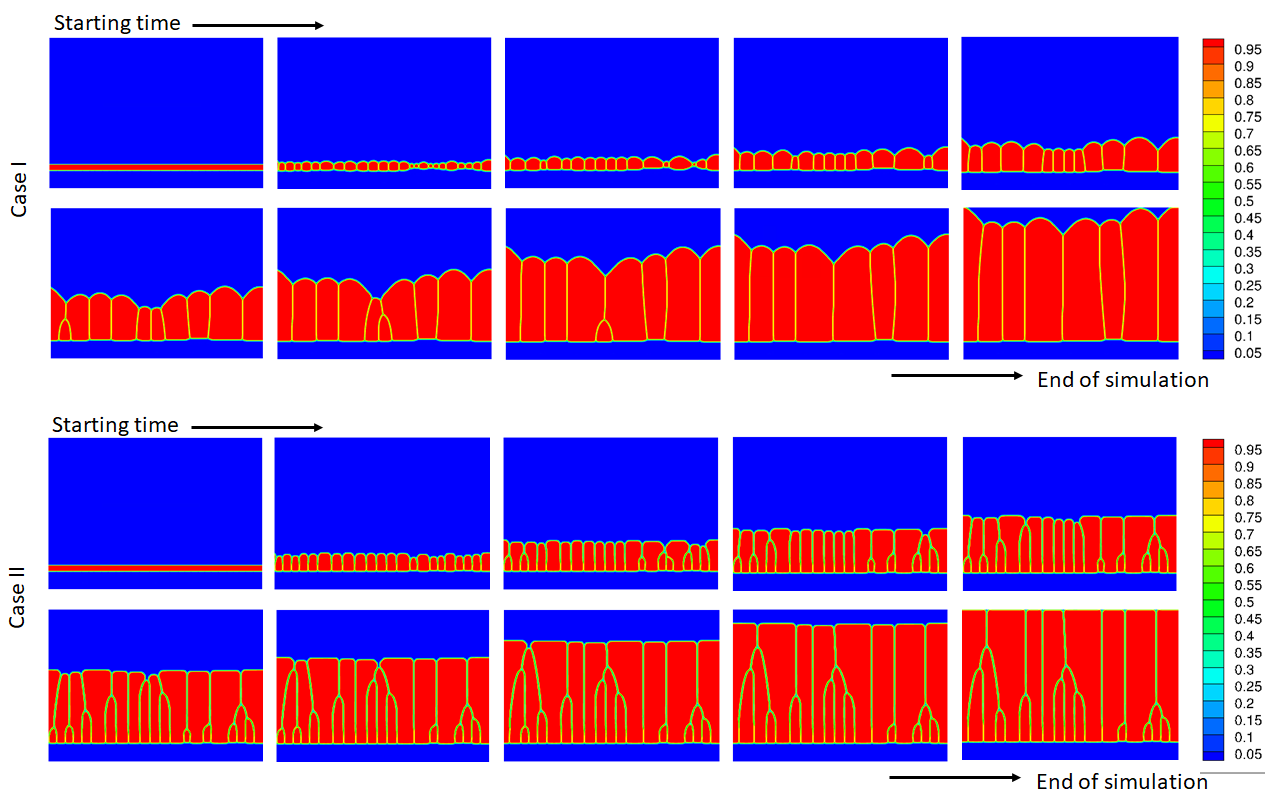}%
    \put(0,61){(a)}%
    \put(0,30){(b)}%
    \end{overpic}
    \begin{overpic}[width=0.99\textwidth]{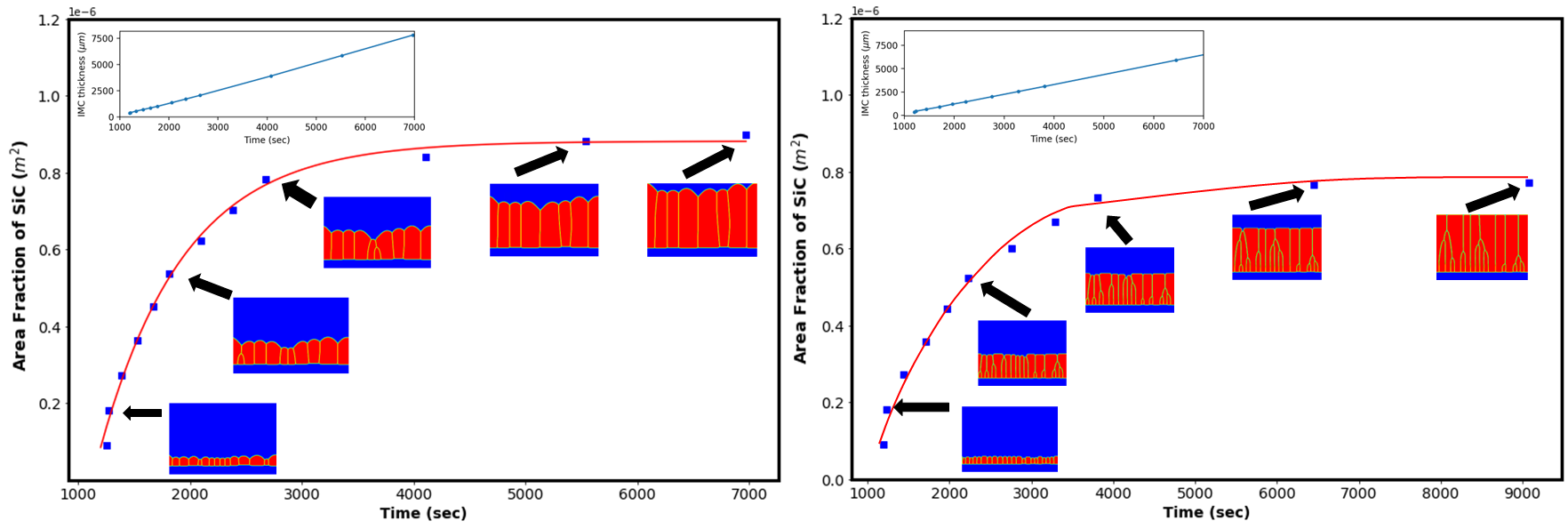}%
    \put(45,29){(c)}%
    \tiny\put(25,5){$A_f=-7.461e^{-07}(e^{-.00131x})+8.824e^{-07}$}%
    \put(95,29){\normalsize{(d)}}%
    \tiny\put(72,5){$A_f=-1.877e^{-06}(e^{-.00089x})+7.761e^{-07}$}%
    \end{overpic}
    \caption{(a) and (b) Morphological evolution of the SiC layer during reactive melt infiltration at T = 1450$^{\circ}$C for two sets of kinetic parameters shown in Table~\ref{tab:pars_two_sims}. (c) and (d) The area faction of the SiC layer with the thickness curve embedded in the top left corner of each graph of these simulations.}
    \label{fig:evolution}
\end{figure*}

 Figures~\ref{fig:evolution}(a) and \ref{fig:evolution}(b) present two instances of SiC layer evolution over time in our simulations, contrasting planar morphological evolution with scallop-like evolution. The color map illustrates the SiC phase-field order parameter profile. Simulation case I, depicted in Figure\ref{fig:evolution}(a), exhibits more grain coalescence. In contrast, simulation case II displays less coalescence but a planar growth. The persistence of smaller grains affects the larger grains that grow around them as SiC extends to the top of the simulation domain, resulting in columnar grain growth due to rapid grain boundary diffusion and lower interface energy between the liquid and SiC. On the other hand, scallop-type grains in simulation case II suggest the possibility of SiC grains dissolving rather than growing completely. The combination of certain parameters allows for either type of growth. Notably, there is minimal planar grain growth observed in our experimental results, indicating that while simulations can initiate planar growth, it may not be feasible experimentally. A comparison of simulation parameters in Table~\ref{tab:pars_two_sims} reveals that the most significant difference lies in the variation of interfacial energies. This underscores that grain boundary and liquid interfacial energies are the key parameters influencing the evolution of the simulations, while diffusivities and mobilities, with similar values, play a relatively minor role in driving these changes.

\par Figures~\ref{fig:evolution}(c) and \ref{fig:evolution}(d) correspond to the simulations that produced the SiC layer evolutions shown in Figures~\ref{fig:evolution}(a) and \ref{fig:evolution}(b). Both figures depict the growth of area fraction over time, along with the evolution of thickness in the upper left-hand corner. As mentioned earlier, the data is fitted to an exponential curve, and equations are provided in these figures. Additionally, simulation evolution images are included to visually represent the progress over time.

\par The post-nucleation initial condition in our simulations imposes limitations related to dissolution or reprecipitation stages of SiC growth. We assume the nucleation stage is "complete" with the formation of a continuous SiC layer at the time we initiate our simulations, and the reaction transitions to being governed by diffusion. It is conceivable that there is a point in the experiment when the SiC layer becomes so thick that C diffusion essentially ceases, indicating a complete reaction in the diffusion-controlled regime. However, in both the conducted experiments and completed simulations, this point was not reached within the specified time frames.


\begin{table*}[!ht]
    \centering
    \begin{center}
    \begin{tabular}{ c c c c } 
     \toprule
     \multirow{2}{*}{Parameter} & Case I & Case II & \multirow{2}{*}{Unit} \\
     
           & \includegraphics[width=0.1\textwidth]{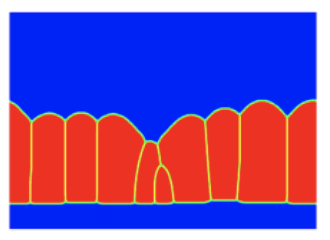} & \includegraphics[width=0.1\textwidth]{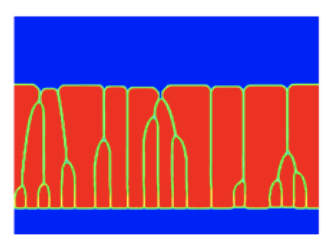} & \\
     \hline
     $D_L$ & $5.40\times10^{-13}$ & $2.97\times10^{-13}$ & $cm^2$/s\\
     $D_S$ & $3.21\times10^{-19}$ & $3.81\times10^{-19}$ & $cm^2$/s\\
     $D_I$ & $2.91\times10^{-18}$ & $6.62\times10^{-19}$ & $cm^2$/s\\
     $D_{IIS}$ & $8.42\times10^{-16}$ & $4.04\times10^{-16}$ & $cm^2$/s\\
     $D_{IIL}$ & $3.40\times10^{-15}$ & $3.01\times10^{-15}$ & $cm^2$/s\\
     $D_{GB}$ & $6.10\times10^{-15}$ & $7.30\times10^{-15}$ & $cm^2$/s\\
     $\sigma_S$ & 0.69 & 0.61 & J/$m^2$\\
     $\sigma_{GB}$ & 0.37 & 0.29 & J/$m^2$\\
     $\sigma_L$ & 0.26 & 0.11 & J/$m^2$\\
     $M_S$ & $9.34\times10^{-08}$ & $9.37\times10^{-08}$ & $m^3/{J.s}$\\
     $M_L$ & $5.76\times10^{-07}$ & $4.52\times10^{-07}$ & $m^3/{J.s}$\\  
     \bottomrule
    \end{tabular}
    \end{center}
    \caption {Parameters with their associated values for the completed simulations in Figure \ref{fig:evolution}. Note, parameters are defined in Table ~\ref{tab:Kinetics}.} \label{tab:pars_two_sims}   
\end{table*}

\par
\begin{figure*}[!ht]
    \centering
    \includegraphics[width=1\textwidth]{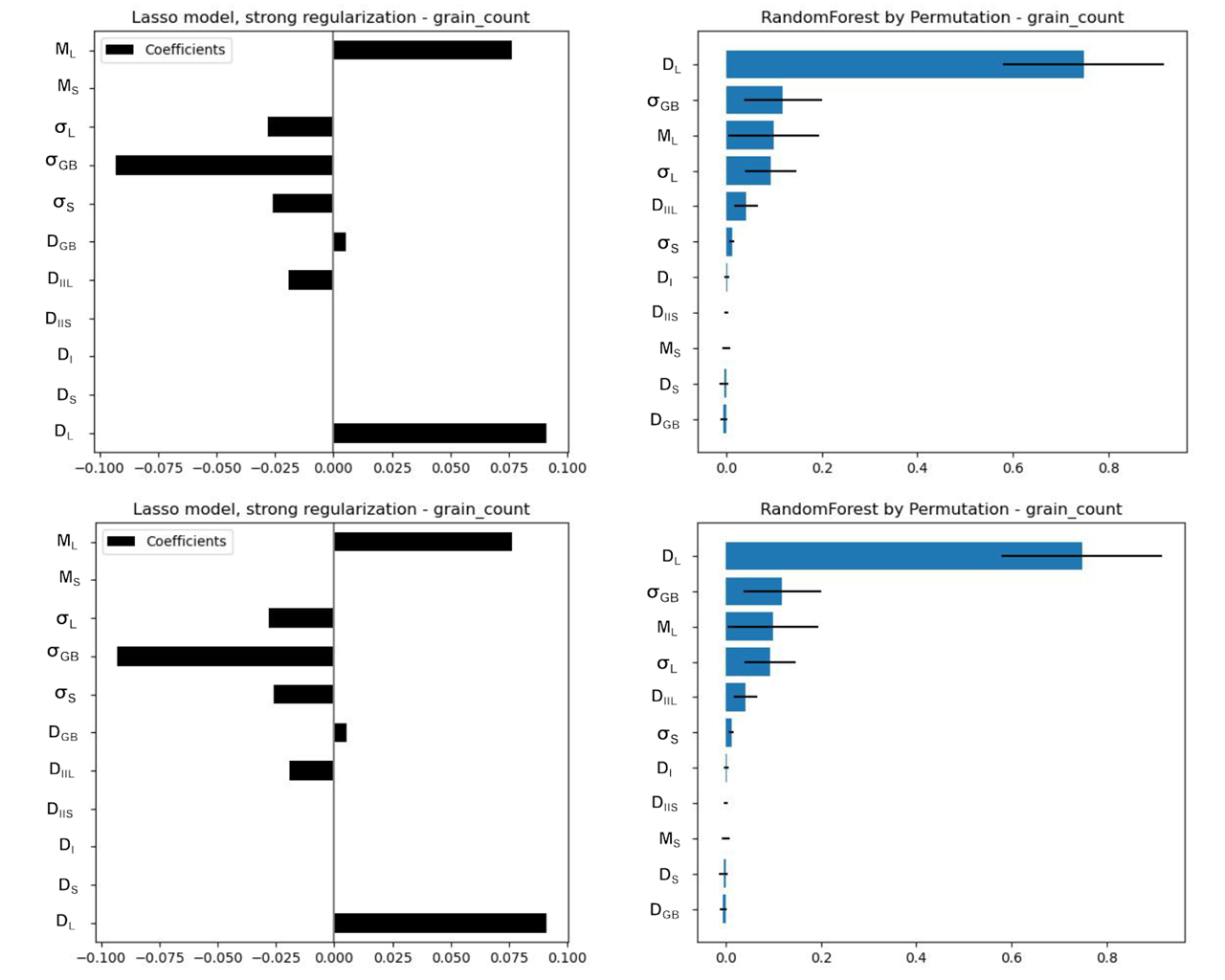}
    \caption{Multi-phase-field model feature importance study for the SiC layer growth (output I) and the SiC grain coalescence (output II). Lasso model coefficient regularization and Random Forest permutation predict a similar rank of importance for liquid diffusion, liquid interface mobility, grain boundary interface energy, and liquid/IMC interface energy for grain count model output. Likewise, these models predict a similar rank of importance for liquid diffusion, liquid/IMC interface diffusion, liquid mobility, and grain boundary interface energy for IMC thickness model output.}
    \label{fig:ML_feature_importance}
\end{figure*}

 We further explored the significance of multi-phase-field model parameters on two key outputs: the average thickness of the SiC layer at a fixed time (Output I) and the grain count in the confined simulation domain after a specific fixed time (Output II). We employed both model-agnostic and model-dependent approaches for feature importance analysis, utilizing Lasso linear regression and permutation feature importance methods. The outcomes depicted in Fig.~\ref{fig:ML_feature_importance} affirm that both machine learning methods identify the same features as the most influential ones for Outputs I and II. Specifically, the diffusion coefficient of liquid Si emerges as the most crucial feature significantly impacting the SiC average layer thickness and grain count over time. Other liquid parameters, such as liquid Si/SiC interface mobility and liquid/SiC interface diffusivity, also hold substantial importance, although the degree of significance may vary depending on the method employed (i.e., Lasso or permutation methods).

\par Regarding SiC layer thickness (Output I), the four most influential model parameters, identified through the permutation by random forest regression ML method, include liquid Si diffusion, Si/SiC interface diffusivity, Si/SiC interface energy, and solid C diffusion. Notably, the first influential parameter is consistent between the two methods. However, the Lasso model diverges slightly, listing Si diffusion, Si/SiC interface mobility, Si/SiC interface energy, and Si/SiC interface diffusivity as the most influential parameters, aligning closely with the permutation method. For grain count (Output II), the four most influential model parameters are liquid Si diffusion, Si/SiC interface diffusivity, SiC/SiC interface energy, and Si/SiC interface energy. Both ML methods precisely agree in this case, and the ranking obtained is consistent for the first five parameters.

\section{Conclusions}\label{sec:conc}

\par Our developed multi-phase-field model offers insights into the kinetics of SiC growth during the reactive melt infiltration process at 1450 $^{\circ}$C. By considering the uncertainty in model parameters arising from limited experimental knowledge in the Si/C system, we explore various SiC morphologies, aligning with concurrent experimental observations. Beyond visualizing the evolving SiC microstructure, our model provides valuable information on energy terms related to grain boundaries and interfaces. This data enables further exploration of the relationship between grain boundary diffusion rates, bulk diffusion, and SiC growth at the interface. Our observations highlight the growth of SiC over the graphite substrate, providing insight on the extent of the kinetic parameters in Si-C system. This investigation serves as a crucial tool for understanding processes at extremely high temperatures, where instrumentation limitations hinder direct kinetic measurements. For experimental validation, we create carbon and SiC powder-based preforms and infiltrate them at different time frames, enabling a comparison between simulated SiC growth and experimental samples. The resulting data not only provides computational values for diffusivities, interfacial energies, and mobilities but also furnishes experimental insights into the kinetics of the reaction. The study demonstrates the qualitative reproduction of experimental results regarding the grain growth of SiC over time. Computational investigation of SiC growth during reactive melt infiltration process proves considerably more intricate than observations from experimental work. SiC grain growth emerges as an 11-dimensional problem with limited quantitative guidance available.

\par The results are proof of the application of the phase field model on the prediction of the microstructure evolution in the Si-C binary system. Both experiments and calculations show similar growth over time. The developed model does have several limitations, including the need to conduct simulations post-nucleation to observe a full picture of SiC evolution. Future refinement of our model will include nucleation to enable the observation of SiC growth from the earliest stages of the melt infiltration process.

\par Our focus extends to identifying strategies for controlling microstructure evolution to align with experimental growth, laying the groundwork for future applications. This knowledge holds promise for ternary systems, offering valuable insights into the utility of this information for other ceramic matrix composites (CMC) processed through reactive melt infiltration. Specifically, applications involving Si-Hf-C or Si-Zr-C systems stand to benefit significantly. Simulations that elucidate the structural evolution in CMC systems contribute to the identification of materials with advantageous properties, holding implications for aerospace, aviation, and nuclear applications.

\section{Appendix}
\par Listed below are the parameters of the energy functions from the SGTE database and others used to obtain energy curves shown in Fig.~\ref{fig:Free_Energy} (equations 9-11). 

\footnotesize

\begin{table*}[!ht]
    \centering
    \small
    \begin{adjustbox}{width=\textwidth,center}
    \begin{tabular}{llc} \toprule
       Parameter & & Reference \\ \midrule
       $G^{Graphite}_C$ & $-17368.441+170.73 \cdot T-24.3 \cdot T \cdot \text{ln(T)}-0.4723E{-3} \cdot T^{2}+2562600 \cdot T^{-1}-264300000 \cdot T^{-2}+1.2E10 \cdot T^{-3}$+Gpres & ~\cite{du1999experimental}
       \\
       $G^{Liq}_C$ & $100000.559+146.1 \cdot T-24.3 \cdot T \cdot \text{ln(T)}-0.4723E{-3} \cdot T^{2}+2562600 \cdot T^{-1}-2.643E{8} \cdot T^{-2} + 1.2E10 \cdot T^{-3}$+Gpres & ~\cite{du1999experimental}
       \\
       $G_{SiC}$ & $-44291.98+135.5731 \cdot T-20.63973 \cdot T \cdot \text{ln(T)}-2.15133E{-3} \cdot T^{2}+4E{5} \cdot T^{-1}+10^{7} \cdot T^{3}$ & ~\cite{du1999experimental}
       \\
       $G^{SER}_{Si}$ & $-9457.642+167.281367 \cdot T-27.196 \cdot T \cdot \text{ln(T)}-420.369E{28} \cdot T^{-9}$ & ~\cite{dinsdale_sgte_1991}
       \\
       $^0L^{Liq}_{ij}$ & $+25644.97-6.38115 \cdot T$ & ~\cite{cupid_thermodynamic_2007} \\
       GHSERSI & $-7.04\times10^{4}$ \\
       GSOLC & $-3.5202\times10^{4}$ \\
       GLIQC &  $3.97\times10^{4}$ \\
       GLIQSI & $-7.15\times10^{4}$ \\
       GSOLSI & $-7.04\times10^{4}$ \\
       GSIC & $-1.63\times10^{5}$ \\
       GCG & $-3.52\times10^{4}$ \\
       GDIAC & $-2.58\times10^{4}$ \\
       GHSERC & $-3.52\times10^{4}$
       \\ \bottomrule
    \end{tabular}
    \end{adjustbox}
    \caption{Thermodynamic parameters used for the calculation of energy curves involved in SiC reactions.}
    \label{tab:my_label}
\end{table*}

\section{Acknowledgement}
\par The authors would like to acknowledge the support from AFRL through sub-contract 165852-19F5830-19-02-C1. MM and EJMIII acknowledge support from the National Science Foundation, grant No. 1545403, Data-Enabled Discovery and Design of Energy Materials, D3EM.  

\label{S:6}

 \bibliographystyle{elsarticle-num-names} 
 \bibliography{Library.bib}





\end{document}